\documentclass{article}
\usepackage{graphicx} 
\usepackage{complexity}
\usepackage{babel}
\usepackage{xspace}
\usepackage{amsthm}
\usepackage[utf8]{inputenc}
\usepackage[T1]{fontenc}
\usepackage{amsmath}
\usepackage{todonotes}
\usepackage{tikz}
\usepackage{authblk}
\usepackage{mathdots}
\usepackage[hidelinks]{hyperref}

\usetikzlibrary{calc, positioning, fit, shapes.misc, shapes.geometric, shapes,arrows}

\def\rotateclockwise#1{
  \newdimen\xrw
  \pgfextractx{\xrw}{#1}
  \newdimen\yrw
  \pgfextracty{\yrw}{#1}
  \pgfpoint{\yrw}{-\xrw}
}

\def\rotatecounterclockwise#1{
  \newdimen\xrcw
  \pgfextractx{\xrcw}{#1}
  \newdimen\yrcw
  \pgfextracty{\yrcw}{#1}
  \pgfpoint{-\yrcw}{\xrcw}
}

\def\outsidespacerpgfclockwise#1#2#3{
  \pgfpointscale{#3}{
    \rotateclockwise{
      \pgfpointnormalised{
        \pgfpointdiff{#1}{#2}}}}
}

\def\outsidespacerpgfcounterclockwise#1#2#3{
  \pgfpointscale{#3}{
    \rotatecounterclockwise{
      \pgfpointnormalised{
        \pgfpointdiff{#1}{#2}}}}
}

\def\outsidepgfclockwise#1#2#3{
  \pgfpointadd{#2}{\outsidespacerpgfclockwise{#1}{#2}{#3}}
}

\def\outsidepgfcounterclockwise#1#2#3{
  \pgfpointadd{#2}{\outsidespacerpgfcounterclockwise{#1}{#2}{#3}}
}

\def\outside#1#2#3{
  ($ (#2) ! #3 ! -90 : (#1) $)
}

\def\cornerpgf#1#2#3#4{
  \pgfextra{
    \pgfmathanglebetweenpoints{#2}{\outsidepgfcounterclockwise{#1}{#2}{#4}}
    \let\anglea\pgfmathresult
    \let\startangle\pgfmathresult

    \pgfmathanglebetweenpoints{#2}{\outsidepgfclockwise{#3}{#2}{#4}}
    \pgfmathparse{\pgfmathresult - \anglea}
    \pgfmathroundto{\pgfmathresult}
    \let\arcangle\pgfmathresult
    \ifthenelse{180=\arcangle \or 180<\arcangle}{
      \pgfmathparse{-360 + \arcangle}}{
      \pgfmathparse{\arcangle}}
    \let\deltaangle\pgfmathresult

    \newdimen\x
    \pgfextractx{\x}{\outsidepgfcounterclockwise{#1}{#2}{#4}}
    \newdimen\y
    \pgfextracty{\y}{\outsidepgfcounterclockwise{#1}{#2}{#4}}
  }
  -- (\x,\y) arc [start angle=\startangle, delta angle=\deltaangle, radius=#4]
}

\def\corner#1#2#3#4{
  \cornerpgf{\pgfpointanchor{#1}{center}}{\pgfpointanchor{#2}{center}}{\pgfpointanchor{#3}{center}}{#4}
}

\def\hedgeiii#1#2#3#4{
  \outside{#1}{#2}{#4} \corner{#1}{#2}{#3}{#4} \corner{#2}{#3}{#1}{#4} \corner{#3}{#1}{#2}{#4} -- cycle
}

\def\hedgeiiii#1#2#3#4#5{
  \outside{#1}{#2}{#5} \corner{#1}{#2}{#3}{#5} \corner{#2}{#3}{#4}{#5} \corner{#3}{#4}{#1}{#5} \corner{#4}{#1}{#2}{#5} -- cycle
}

\def\hedgem#1#2#3#4{
  
  \outside{#1}{#2}{#4}
  \pgfextra{
    \def\hgnodea{#1}
    \def\hgnodeb{#2}
  }
  foreach \c in {#3} {
    \corner{\hgnodea}{\hgnodeb}{\c}{#4}
    \pgfextra{
      \global\let\hgnodea\hgnodeb
      \global\let\hgnodeb\c
    }
  }
  \corner{\hgnodea}{\hgnodeb}{#1}{#4}
  \corner{\hgnodeb}{#1}{#2}{#4}
  -- cycle
}

\usepackage{colortbl}
\definecolor{roug}{RGB}{153,0,0}
\definecolor{fondt}{RGB}{255,255,240}
\colorlet{fond}{fondt!40}
\definecolor{orang}{HTML}{FF851B}
\definecolor{orangfonce}{RGB}{255,70,0}
\definecolor{coul3}{RGB}{255,204,0}
\definecolor{coul5}{RGB}{255,102,102}
\definecolor{violetarete}{RGB}{118,55,96}
\definecolor{joli_bleu}{RGB}{41,128,185}
\definecolor{joli_rouge}{RGB}{192, 57, 43}
\definecolor{joli_vert}{RGB}{46, 204, 113}
\definecolor{joli_mauve}{RGB}{155, 89, 182}
\definecolor{joli_orange}{RGB}{243, 156, 18}
\definecolor{bleujoli}{HTML}{17C0EB}
\definecolor{rosejoli}{RGB}{176.,0,64.}
\definecolor{vertjoli}{RGB}{34, 120, 15}
\definecolor{rougejoli}{RGB}{217.,1,21.}
\definecolor{mauvejoli}{HTML}{9B59B6}

\def\tikzbox#1#2#3#4{
\begin{center}
\begin{tikzpicture}
\tikzstyle{rec} = [rectangle,rounded corners, thick,draw=#1!90,fill=fond] 
\path (0,0) node[rec,text=black,inner sep=3mm] (theoreme){\parbox{0.95\textwidth}{#4}};
\node[rec,text=#1!95,anchor=south west,xshift=3mm,yshift=-1mm] at (theoreme.north west) {\textbf{#2} #3};
\end{tikzpicture}
\end{center}
\vspace{-0.2cm}
}

\def\mbox#1#2{\tikzbox{mauvejoli}{}{#1}{#2}}

\tikzstyle{noeud}=[circle, inner sep=1.5, minimum size =2 pt, line width = .8pt, draw=black, fill=white]
\tikzstyle{R}=[circle, inner sep=1.5, minimum size =2 pt, line width = 1.5pt, draw=rougejoli, fill=rougejoli]
\tikzstyle{B}=[circle, inner sep=1.5, minimum size =2 pt, line width = 1.5pt, draw=bleujoli, fill=bleujoli]
\tikzstyle{inv}=[circle,inner sep=0, minimum size =4 pt, line width = 1pt, draw=white, fill=white, text= black]

 \tikzstyle{rond}=[circle, draw = red, inner sep=3mm]
 \tikzstyle{croix}=[cross out, draw = blue, inner sep=3mm]

\tikzstyle{v}=[circle,inner sep=0, minimum size =4 pt, line width = 1pt, draw=black, fill=black, text= white]

\tikzstyle{ghost}=[circle,inner sep=0, minimum size =4 pt, line width = 1pt, draw=black!20, fill=black!20, text= white]

\tikzstyle{arete}=[draw, line width=1.5pt, draw=black]

\tikzstyle{decision} = [diamond, draw, fill=yellow!20, 
    text width=10em, text centered, node distance=5cm, inner sep=0pt]
\tikzstyle{block} = [rectangle, draw, fill=yellow!20, 
    text width=10em, text centered, rounded corners, minimum height=4em]
\tikzstyle{line} = [draw, -latex']
    
\tikzstyle{sortieD} = [draw, ellipse,fill=blue!20]
 \tikzstyle{sortieA} = [draw, ellipse,fill=red!20]

\newcommand{\QBF}{\lang{QBF}\xspace}
\newcommand{\strat}{\mathcal{S}}
\newtheorem{theorem}{Theorem}
\newtheorem{lemma}[theorem]{Lemma}

\newtheorem{corollary}[theorem]{Corollary}

\newcommand{\hxf}{\hyp = (\som, \WS)}
\newcommand{\hxfp}{\hyp' = (\som', \WS')}
\newcommand{\hyp}{\mathcal{H}}
\newcommand{\WS}{\mathcal{F}}
\newcommand{\som}{\mathcal{X}}
\newcommand{\etal}{\textit{et al.} }

\title{Bounded degree QBF and positional games }
\author{Nacim Oijid}
\date{}
\affil{Univ. Bordeaux, CNRS,  Bordeaux INP, LaBRI, UMR 5800, F-33400, Talence, France}
\affil{This research was partly supported by the ANR project P-GASE (ANR-21-CE48-0001-01)}

\begin{document}

\maketitle

\begin{abstract}
The study of \SAT\ and its variants has provided numerous \NP-complete problems, from which most \NP-hardness results were derived. Due to the \NP-hardness of \SAT, adding constraints to either specify a more precise \NP-complete problem or to obtain a tractable one helps better understand the complexity class of several problems. In 1984, Tovey proved that bounded-degree \SAT\ is also \NP-complete, thereby providing a tool for performing \NP-hardness reductions even with bounded parameters, when the size of the reduction gadget is a function of the variable degree. In this work, we initiate a similar study for \QBF, the quantified version of \SAT. We prove that, like \SAT, the truth value of a maximum degree two quantified formula is polynomial-time computable. However, surprisingly, while the truth value of a 3-regular 3-\SAT\ formula can be decided in polynomial time, it is \PSPACE-complete for a 3-regular \QBF\ formula. A direct consequence of these results is that Avoider-Enforcer and Client-Waiter positional games are \PSPACE-complete when restricted to bounded-degree hypergraphs. To complete the study, we also show that Maker-Breaker and Maker-Maker positional games are \PSPACE-complete for bounded-degree hypergraphs.
\end{abstract}

\section{Introduction}

The \NP-completeness of \SAT\ established by Cook~\cite{Cook1971} marked the beginning of the study of computational hardness. To better understand these problems and explore the boundary between tractable and \NP-hard problems, several variants of \SAT\ have been introduced, providing proofs that many problems are \NP-complete, even when restricted to specific instances. The most commonly studied variant is $3$-\SAT, where \SAT\ is restricted to clauses of size at most $3$. However, other variants also appear frequently, depending on the context, such as planar $3$-\SAT\ or $1$-in-$3$-\SAT. For examples of such variants and reductions requiring specific conditions on the formulas, we refer the reader to Karp's article~\cite{Karp1972} and Papadimitriou's book~\cite{papadimitriou1994}.

In \PSPACE, the central problem from which most reductions are derived is $3$-\QBF, the quantified version of $3$-\SAT, i.e., quantified Boolean formulas restricted to clauses of size at most $3$. In this paper, we adopt the two-player game interpretation of \QBF\ proposed by Stockmeyer and Meyer~\cite{stockmeyer1973} when they proved its \PSPACE-completeness. Let $\psi = Q_1 x_1, \dots, Q_n x_n \varphi$ be a quantified Boolean formula, where for each $1 \leq i \leq n$, $Q_i \in \{\exists, \forall\}$ and $\varphi$ is a quantifier-free Boolean formula over $x_1, \dots, x_n$. The \QBF\ game played on $\psi$ involves two players, Satisfier and Falsifier, as follows: for $i = 1$ to $n$, if $Q_i = \exists$, Satisfier assigns a truth value to $x_i$, and if $Q_i = \forall$, Falsifier assigns a truth value to $x_i$. Once all variables have been assigned values, a valuation $\nu$ for $x_1, \dots, x_n$ is obtained, and Satisfier wins if and only if $\nu$ satisfies $\varphi$.

Let $\varphi$ be a Boolean formula, and let $x$ be a variable in $\varphi$. We denote by $d(x)$ the \emph{degree} of $x$ in $\varphi$, i.e., the number of clauses in which $x$ appears. We also denote by $\Delta(\varphi)$ the \emph{degree} of $\varphi$, which is the maximum degree of any variable in $\varphi$. A formula $\varphi$ is said to be \emph{$k$-regular} if every variable in $\varphi$ has degree $k$. If $c \in \varphi$ is a clause, we denote by $|c|$ its \emph{size}, which is the number of literals in $c$. The \emph{rank} of a formula $\varphi$ is the largest size of any clause in $\varphi$. A formula $\varphi$ is said to be \emph{$k$-uniform} if all its clauses have size exactly $k$.

This paper focuses on bounded-degree formulas, i.e., formulas in which each variable appears in a bounded number of clauses. Bounded-degree \SAT\ has been studied since 1984, when Tovey~\cite{tovey1984} proved that \SAT\ remains \NP-complete when restricted to formulas with a maximum degree of $3$, but is polynomial-time solvable for formulas with a maximum degree of $2$ or for $3$-regular $3$-uniform formulas.

We initiate here the study of bounded-degree \QBF. For integers $r, d$, we denote by $r$-\QBF-$d$ the set of Boolean formulas in which the clauses have size at most $r$ and where each variable appears in at most $d$ clauses. If either $r$ or $d$ is unspecified, there are no restrictions on them. This study is motivated by the fact that \QBF\ is a central problem for \PSPACE\ hardness reductions, and restricted versions of it can lead to \PSPACE-hard problems on specific instances. The motivation for this study primarily stems from the desire to understand bounded-degree positional games. Positional games were introduced by Hales and Jewett~\cite{Hales1963} in 1963, and the study of their complexity began in 1978 when Schaefer proved that the Maker-Breaker version is \PSPACE-complete~\cite{schaefer1978}. In this convention, two players, Maker and Breaker, alternate claiming the vertices of some hypergraph $\hyp$. Maker wins if the vertices she manage to claim all the vertices of some hyperedge of $\hyp$, otherwise, Breaker wins. More recently, other conventions have been proven to be \PSPACE-complete, including Maker-Maker~\cite{byskov2004}, Avoider-Avoider~\cite{Fenner2015, Burke2019}, Avoider-Enforcer~\cite{Gledel2023}, and Client-Waiter~\cite{gledel2024}. However, all these proofs focused on the rank of the hypergraph (i.e., the size of its largest hyperedge) and did not attempt to optimize its maximum degree. For a general overview of the complexity of positional games, we refer the reader to Oijid's Ph.D. thesis~\cite{oijid2024thesis}.

Recent studies on positional games focus on games played on graphs, where certain graph structures are considered as the winning sets~\cite{ Slany2000, Duchene2020, duchêne2023complexitymakerbreakergamesedge, Bensmail2023}. In most of these studies, both the rank and the degree of the vertices of the hypergraph play important roles in the reductions provided. Bounding the degree of vertices, in conjunction with the rank, would enable reductions to be restricted to bounded-degree graphs. This result is particularly relevant within the parameterized complexity framework, as bounded-degree graphs have locally bounded treewidth. Bonnet \etal~\cite{Bonnet2017} showed that positional games parameterized by the number of moves are \FPT\ on locally bounded treewidth graphs, which indicates that tractability really requires the number of moves to be treated as a parameter.

In this paper, we prove in Section~\ref{sec:QBF} that $2$-\QBF\ can be solved in quadratic time, and that, unlike the unquantified version, $3$-\QBF-$3$ is already \PSPACE-complete. In Section~\ref{sec:pogames}, we explore the computational complexity of positional games restricted to bounded-degree hypergraphs. As a direct consequence of the \PSPACE-hardness of $3$-\QBF-$3$, we prove that Avoider-Enforcer and Client-Waiter games are \PSPACE-complete. However, since the reduction for Maker-Breaker games generates vertices of unbounded degree, we prove separately that Maker-Breaker games are \PSPACE-complete when restricted to hypergraphs of rank 12 and maximum degree 5. As a consequence, we answer an open question from Bagan et.al.~\cite{Bagan2024}, proving that the Maker-Breaker Domination game is \PSPACE-complete, even restricted to bounded degree graphs, and we prove that Maker-Maker games are also \PSPACE-complete restricted to bounded degree hypergraphs.

\section{Complexity of bounded degree \QBF}\label{sec:QBF}

\subsection{\QBF-2 is Polynomial}

This section focuses on quantified formulas where each variable appears in at most two clauses. Tovey proved in 1984~\cite{tovey1984} that determining whether a \SAT\ formula where the variables appear each in at most two clauses is polynomial time solvable. However, his proof relies on a lemma stating that a \SAT\ formula in which all the clauses have size at least two and each of the variables appears exactly once complemented and once uncomplemented is True. This lemma cannot be applied to universal quantified variables.

\begin{theorem}
    The winner of the \QBF game played on a \QBF-2 formula can be computed in quadratic time.
\end{theorem}

\begin{proof}
    Let $\psi = Q_1 x_1, \dots, Q_n x_n \varphi$ be a \QBF\ formula in which all the variables have degree at most $2$.

We consider the following reduction rules:

\begin{enumerate}
    \item If a clause $C$ contains both one literal and its complement, remove $C$.
    \item If $Q_n = \forall$, remove $x_n$ from the clauses containing it (either complemented or not). If this step makes a clause empty, $\psi$ is False.
    \item If $Q_n = \exists$, and if $x_n$ appears only uncomplemented or only complemented, remove the clauses containing $x_n$ and $x_n$.
    \item If $Q_n = \exists$, and if $x_n$ appears once uncomplemented and once complemented, let $C_i = \ell_i^1 \vee \dots \vee \ell_i^k \vee x_n$ and $C_j = \ell_j^1 \vee \dots \vee \ell_j^{k'} \vee \neg x_n$  the two clauses containing $x_n$. Replace in $\varphi$ $C_i$ and $C_j$ by $C' = \ell_i^1 \vee \dots \vee \ell_i^k \vee \ell_j^1 \vee \dots \vee \ell_j^{k'}$, and remove $x_n$.
\end{enumerate}

Once no reduction rule can be applied, if the process did not stop because of $(2)$, then $\psi$ has no clause left, and therefore is True. 

Each of this reduction can be applied in linear time and each of them decreases the number of variables. Thus, applying reductions until a truth value is returned can be done in quadratic time. It remains to prove that each reduction rule does not change the outcome of the game.

\begin{enumerate}
    \item If a clause contains both a literal and its complement, it will always be satisfied by the variable corresponding to this literal. Thus, it can be removed without changing the outcome of the game.
    \item Suppose $Q_n = \forall$. First, if $x_n$ is the only one variable in a clause $C_0$, then $\psi$ is False, since either putting $x_n$ to True or to False will evaluate $C_0$ to False and Falsifier can always perform this move. Now, let $\psi'$ be the formula, in which $x_n$ has been removed from the clauses containing it. We prove that $\psi$ and $\psi'$ have the same outcome. Let $C$ and $C'$ be the two clauses containing either $x_n$ or $\neg x_n$ we can have $C = C'$ if $x_n$ is in a single clause. Suppose first that $\psi$ is Satisfier has a winning strategy on $\psi$. Since $C$ and $C'$ cannot contain both $x_n$ and $\neg x_n$ otherwise they would have been handled by $(1)$, and since $x_n$ is the last variable to be played, both $C$ and $C'$ are still satisfied when Falsifier plays $x_n$, otherwise a move of Falsifier would make him win. Thus, they are satisfied by another variable. Thus Satisfie wins of $\psi'$. Reciprocally, if Satisfier wins on $\psi'$, she also wins on $\psi$ since all its clauses are in $\psi'$ and $C$ and $C'$ have a subset of their literals as clauses of $\psi'$.
    \item Suppose $Q_n = \exists$ and $x_n$ only appears complemented or uncomplemented. Since $x_n$ will be played by Satisfier, and since $x_n$ either always appears complemented or uncomplemented, Satisfier can choose its value, regardless the value of the other variables. Thus, by choosing the value that will satisfy all the clause in which $x_n$ is, she ensure to satisfy all of them.
    \item Suppose $Q_n = \exists$. Let $C_i = \ell_i^1 \vee \dots \vee \ell_i^k \vee x_n$ be the clause where $x_n$ appears uncomplemented and $C_j = \ell_j^1 \vee \dots \vee \ell_j^{k'} \vee \neg x_n$ be the clause where $x_n$ appears complemented. Let $\psi'$ be the formula obtained by replacing $C_i$ and $C_j$ in $\psi$ by $C' = \ell_i^1 \vee \dots \vee \ell_i^k \vee \ell_j^1 \vee \dots \vee \ell_j^{k'}$. 
    
    Suppose Satisfier wins on $\psi$. Let $\nu$ be a valuation of $x_1, \dots x_{n-1}$ chosen according to the moves of Satisfier and falsifier such that $\nu$ such that Satisfier plays according his winning strategy on $\psi$. In particular either $C_i$ or $C_j$ will not be satisfied by the chosen valuation of $x_i$. Thus, it is satisfied by another literal, which is also in $C'$. Thus, the same strategy make her win on $\psi'$.

    Suppose now that Satisfier wins on $\psi'$. Let $\nu$ be a valuation of $x_1, \dots, x_{n-1}$ following her winning strategy on $\psi'$. Since $\psi'$ is satisfied, there exists an integer $k$ such that $\nu(\ell_i^k) = \top$ or $\nu(\ell_j^k) = \top$. If this literal is a literal $\ell_i^k$, Satisfier sets $\nu(x_n) = \bot$, and in $\psi$, $C_i$ is satisfied by $\ell_i^k$ and $C_j$ is satisfied by $\neg x_n$. Otherwise, she sets $\nu(x_n) = \top$, and in $\psi$, $C_j$ is satisfied by $\ell_j^k$ and $C_i$ is satisfied by $x_n$. Thus, she wins on $\psi$. \qedhere
\end{enumerate}
\end{proof}

\subsection{3-\QBF-3 is \PSPACE-complete}

Tovey~\cite{tovey1984} proved in 1984 that any $3$-\SAT\ formulas in which all the variables appear in exactly three clauses is satisfiable, and therefore their truth value is computed in constant time (linear if you first check that the formula satisfies the hypothesis). In this section, we look at the same problem but in a quantified point of view. Clearly, they cannot be always satisfiable since if all the vertices are quantified by universal quantifiers, it is easy to see that the formula cannot be satisfied unless it is at trivial one. We prove here that computing the winner of the $3$-\QBF-$3$ game is already \PSPACE-complete showing a larger gap between quantified and unquantified formulas

\begin{theorem}\label{3-QBF-3 PSPACE complete}
    Determining the truth value of a $3$-\QBF-$3$ formula is \PSPACE-complete, even restricted to formulas where each clause has size exactly tree and each variable appears in exactly 3 clauses.
\end{theorem}

\begin{proof}
    We provide a reduction from $3$-\QBF. Let $\psi = Q_1 x_1, \dots, Q_n x_n \varphi$ be a $3$-\QBF\  formula. Up to remove some of the variables, we can suppose that each of them appears in at least one clause. Up to add one or two universal variables that only appears positively, we can suppose that each clause has size exactly three, and up to add two new copies of each clauses, we can suppose that each variable appears in a number of clauses which is a multiple of $3$. We construct a $3$-\QBF-$3$ $\psi'$ formula from $\psi$ as follows: For $i = 1$ to $n$, Let $C^i_1, \dots C^i_{3k}$ be the clauses where $x_i$ appears in $\psi$. We transform $Q_i x_i$ to $Q_i x^i_1 \exists x^i_2 \exists x^i_3 \dots \exists x^i_{3k} \forall y^i_1 \dots \forall y^i_k$. In each clause $C^i_j$ for $1 \le j \le 3k$, we replace $x_i$ by $x^i_j$ and for $1 \le j \le 3k$, we add the clauses $x^i_j \vee \neg x^i_{j+1} \vee y^i_{\left \lceil \frac{j}{3} \right \rceil}$ (with $x^i_{3k+1} = x^i_1$).  By construction, all the new clauses have size $3$ and all the variables appear in exactly three clauses. Therefore, $\psi'$ is a $3$-\QBF-$3$ formula. Now, we prove that the $\QBF$ game played on $\psi$ and $\psi'$ have the same outcome.

From this construction, since the variables $y^i_j$ only appears positively and are to be played by Falsifier, it is always optimal for him to put them to $\bot$. It follows that whenever a valuation is given to $x^i_1$, the same valuation has to be chosen for all the $x^i_j$, in order to satisfy all the clauses $x^i_j \vee \neg x^i_{j+1} \vee y^i_{\left \lceil \frac{j}{3} \right \rceil}$, which is possible since Satisfier chooses the valuation of all the $x^i_j$s with $j \ge 2$. Thus the player that have a winning strategy on $\psi$ can apply it on $\psi'$, which proves that the two formulas have the same outcome.
\end{proof}

\section{Complexity of bounded degree positional games are \PSPACE-completes}\label{sec:pogames}

In this section, we prove that four of the most studied conventions of positional games are still \PSPACE-complete when restricted to bounded degree hypergraphs. While the proof is quite immediate for Avoider-Enforcer and Client-Waiter games, since it will be sufficient to count how the degree evolve in the already known reductions. This method will not be sufficient for Maker-Breaker and Maker-Maker games, and we will have to prove it by a new reduction.

\subsection{Avoider-Enforcer and Client-Waiter games}

The proofs that Avoider-Enforcer games and Client-Waiter gales are \PSPACE-complete are reductions from \QBF, with an intermediate problems for Client-Waiter. This reductions only adds a constant number of hyperedges per clause in which a variable is, and thus can be used, together with Theorem~\ref{3-QBF-3 PSPACE complete}, to prove that these games are still \PSPACE-complete restricted to bounded degree  hypergraphs.

\begin{theorem}
    Computing the winner of an Avoider-Enforcer  game is \PSPACE-complete, even restricted to hypergraphs of rank~$6$ and maximum degree~$8$
\end{theorem}

\begin{proof}
    The construction provided by Gledel and Oijid~\cite{Gledel2023} is a reduction from \QBF. Let $\psi$ be a $3$-\QBF-$3$ formula on $2n$ variables and $m$ clauses. Up to add variables in no clauses, we can suppose that the quantifiers of $\psi$ alternates between $\exists$ and $\forall$.
    
    The reductions of Gledel and Oijid creates a hypergraph $\hyp$ on $10n$ vertices  $x_1, \overline{x_1}, \dots, x_{2n}, \overline{x_{2n}}, u_1, u_2, \dots, u_{6n}$.
For each variable $X_i$ of $\psi$ the following $8$ hyperedges are added:

    \begin{minipage}{.46\textwidth}
\begin{align*}
    A_{2i} &=  \hspace{.1cm} ( x_{2i}, \overline{x_{2i}}, u_{6i+1}, u_{6i+3} ) &\\
    C^+_{6i} &=  \hspace{.1cm} ( u_{6i}, u_{6i+1}, u_{6i+3}, x_{2i} ) &\\
    C^+_{6i-2} &=  \hspace{.1cm} ( u_{6i-2}, u_{6i-1}, u_{6i+1}, x_{2i} ) &\\
    C^+_{6i-4} &=  \hspace{.1cm} ( u_{6i-4}, u_{6i-3}, u_{6i-1}, x_{2i-1} ) &
\end{align*}
\end{minipage}
\begin{minipage}{.46\textwidth}
\begin{align*}
    B_{2i-1} &=  \hspace{.1cm} ( x_{2i-1}, \overline{x_{2i-1}}, u_{6i-1} ) &\\
    C^-_{6i} &=  \hspace{.1cm} ( u_{6i}, u_{6i+1}, u_{6i+3}, \overline{x_{2i}} ) &\\
    C^-_{6i-2} &=  \hspace{.1cm} ( u_{6i-2}, u_{6i-1}, u_{6i+1}, \overline{x_{2i}} ) &\\
    C^-_{6i-4} &=  \hspace{.1cm} ( u_{6i-4}, u_{6i-3}, u_{6i-1}, \overline{x_{2i-1}} ) & 
\end{align*}
\end{minipage}

\medskip

Moreover, for each clause $F_j = l^j_{1} \vee l^j_{2} \vee l^j_{3} \in \psi$ where $l^j_{1}, l^j_{2}$ and $l^j_{3}$ are literals either positive or negative, a hyperedge $D_j$ is added. For $k = 1,2,3$, if $l^j_{k}$ is a positive variable $X_p$, then $x_p$ is in $D_j$, if $l^j_{k}$ is a negative one $\neg X_p$, then $\overline{x_p}$ is in $D_j$. Moreover, If $p$ is odd, then $u_{6p -1}$ is in $D_j$, if $p$ is even, then $u_{6p +1}$ is in $D_j$.

We can now count the degree of the vertices in this hypergraph. Let $0 \le i \le n$ be an integer
\begin{itemize}
    \item $d(u_{2i}) = 2$, since this vertex only appears in $C_{2i}^+$ and $C_{2i}^-$.
    \item $d(u_{6i+1}) \le 8$, since this vertex appear in $A_{2i}$, $C_{6i}^+$, $C_{6i}^-$, $C_{6i-2}^+$, $C_{6i-2}^-$ and at most three times in clauses $D_j$.
    \item $d(u_{6i+3}) \le 8$, since this vertex appear in $A_{2i}$, $C_{6i}^+$, $C_{6i}^-$, $C_{6i+2}^+$, $C_{6i+2}^-$ and at most three times in clauses $D_j$.
    \item $d(u_{6i+5)} \le 8$, since this vertex appear in $B_{2i+1}$, $C_{6i-2}^+$, $C_{6i-2}^-$, $C_{6i-4}^+$, $C_{6i-4}^-$ and at most three times in clauses $D_j$.
    \item $d(x_{2i}) \le 6$, since this vertex appear in $A_{2i}$, $C_{6i}^+$, $C_{6i-2}^+$, and at most three times in clauses $D_j$.
    \item $d(\overline{x_{2i}}) \le 6$, since this vertex appear in $A_{2i}$, $C_{6i}^-$, $C_{6i-2}^-$, and at most three times in clauses $D_j$.
    \item $d(x_{2i-1}) \le 5$, since this vertex appear in $B_{2i}$, $C_{6i-4}^+$, and at most three times in clauses $D_j$.
    \item $d(\overline{x_{2i-1}}) \le 5$, since this vertex appear in $B_{2i}$, $C_{6i-4}^-$, and at most three times in clauses $D_j$.
\end{itemize}

Finally, all the vertices of $\hyp$ have degree at most $8$.
\end{proof}

To obtain a similar result for Client-Waiter games, we first need to bound the degree in {\sc Paired SAT}, which was introduced by Gledel, Oijid, Tavenas and Thomassé~\cite{gledel2024}, from which the reduction is made for Client-Waiter games. The {\sc Paired SAT} game is played and $(\varphi, X)$ where $X$ is a set of pair of variables and $\varphi$ is a formula over $X$. The game is played as follows: while there is an unselected pair of variables in $X$, Satisfier selects an unselected pair of variable $(x,y)$ and gives a valuation to $x$. Then Falsifier gives a valuation to $y$. When all the pairs have been selected, Satisfier wins if the valuation created satisfies $\varphi$, otherwhise Falsifier wins.

\begin{lemma}
    {\sc Paired SAT} is \PSPACE-complete, even restricted to formula of maximum degree~7.
\end{lemma}

\begin{proof}
    We consider the reduction provided by Gledel \etal~\cite{gledel2024}

Let $\psi$ be a $3$-\QBF-$3$ formula. Up to add variables in no clauses, we can suppose $\psi = \exists x_1 \forall y_1 \cdots \exists x_n \forall y_n, \varphi$. They constructed an instance of the {\sc Paired-SAT}-game $(\varphi', X)$ as follows: 

\begin{itemize}
    \item $X = \{(z_0, y_0), (x_1, t_1), (z_1, y_1), \dots, (x_n, t_n), (z_n, y_n)\}$, where $y_0$, the $z_i$s, and the $t_j$s are new variables.
    \item $\varphi' = \varphi \wedge \bigwedge\limits_{1 \le i \le n} (y_{i-1} \oplus t_i \oplus z_i)$. 
    
    where $\oplus$ refers to the XOR operator:
    \[a \oplus b \oplus c = (a \vee b \vee c) \wedge (\neg a \vee \neg b \vee c) \wedge (\neg a \vee b \vee \neg c) \wedge (a \vee \neg b \vee \neg c).\]
\end{itemize}

Considering this construction, since all the variables of the $3$-\QBF-$3$ formula appear in at most three clauses, the vertices of the {\sc Paired SAT} formula appear each in at most $7$ clauses. Thus {\sc Paired SAT} is \PSPACE-complete, even restricted to formulas of maximum degree $7$.
\end{proof}

\begin{theorem}
    Computing the winner of a Client-Waiter game is \PSPACE-complete, even restricted to hypergraphs of maximum degree~$35$ and rank~$6$.
\end{theorem}

\begin{proof}
    Let $(\varphi, X)$ be an instance of the {\sc Paired SAT} game of maximum degree $7$. Without loss of generality, we can suppose that there is no clause containing only variables that will be chosen by Falsifier, otherwise, Falsifier can win by putting all the literals of these clause to $\bot$.
    The reduction to Client-Waiter games creates the following hypergraph $\hxf$:

$\som$ is a set of $8n$ vertices: $\som = \bigcup_{1 \le i \le n} S_i \cup F_i $, with for $1 \le i \le n$, $S_i = \{s_{i}^0,s_{i}^T,s_{i}^F,s_{i}^1\}$ and $F_i = \{f_{i}^0,f_{i}^T,f_{i}^{T'},f_{i}^F\}$.

The hyperedges are constructed as follows: 

\begin{itemize}
    \item The block-hyperedges $B = \bigcup_{1 \le i \le n} B_i$:
\[
    B_i = \left\{ H \subseteq S_i \mid  \lvert H \rvert=3\} \cup \{H \subseteq F_i \mid  \lvert H \rvert=3 \right\}.
\]

\item The pair-hyperedges $P = \bigcup_{1 \le i \le n} P_i$:
\[ 
    P_i = \left\{ \{s_i^0, s_i^T, f_i^0, f_i^T \}, \{s_i^0, f_i^F, f_i^T, s_i^{F} \}, \{s_i^0, f_i^F, s_i^T, f_i^{T'} \}, \{s_i^0, s_i^F, f_i^0, f_i^{T'} \} \right \}.
\]

\item The clause-hyperedges. Each clause $C_j \in \varphi$ is a set of three literals $\{\ell_{j}^1,\ell_{j}^2,\ell_{j}^3\}$. They define first, for $1 \le j \le m$ and $k \in \{ 1, 2, 3\}$, the set $H_{j}^k$. 
\begin{align*}
    H_{j}^k = \begin{cases} 
    \{\{s_{i}^0,s_{i}^T\}\} & \text{if } \ell_{j}^k = x_i \\
    \{\{s_{i}^0,s_{i}^F\}\} & \text{if } \ell_{j}^k = \neg x_i \\
    \{\{f_{i}^0,f_{i}^T\},\{f_{i}^0,f_{i}^{T'}\}\} & \text{if } \ell_{j}^k =  y_i \\
    \{\{f_{i}^F\}\} & \text{if } \ell_{j}^k = \neg y_i. 
    \end{cases}
\end{align*}

They then define the set of hyperedges:
   $$ C = \bigcup_{C_j \in \varphi} H_j. $$

\noindent with $H_j = \left\{ h_{1} \cup h_{2} \cup h_{3} \mid \forall k \in \{ 1,2,3 \},\, h_k \in H_{j}^k\right\}$

For example, if $C_j = x_1 \vee y_1 \vee \neg y_2$, we have $H_j^1 = \{ \{s_1^0, s_1^T\} \}$, $H_j^2 = \{ \{f_1^0, f_1^T\}, \{ f_1^0, f_1^{T'}\} \}$ and $H_j^3 = \{\{ f_2^F\}\}$. Finally, we have two hyperedges to encode $C_j$: $H_j = \{ (s_1^0, s_1^T, f_1^0, f_1^{T}, f_2^F), (s_1^0, s_1^T, f_1^0, f_1^{T'}, f_2^F)  \}$

\end{itemize}

Let $x \in \som$ be a vertex. $x$ is in at most $3$ block hyperedge, $4$ pair-hyperedge, and $28$ clause-hyperedge (since at all the clauses will contain at least one vertex $s_i^0$, and therefore at most four clause-hyperedge are created per clause of $\varphi$. Thus $d(x) \le 35$, which ends the proof.
\end{proof}

\subsection{Maker-Breaker games}

Contrary to the reductions for Avoider-Enforcer games and Client-Waiter games, the reduction provided by Schaefer~\cite{schaefer1978} Rahman and Watson~\cite{rahman2021} for Maker-Breaker games creates vertices of arbitrarily large degree. Therefore, we cannot just adapt the proof, and we need to prove their \PSPACE-hardness on bounded degree hypergraphs directly.

\begin{theorem}\label{thm: MB deg 5 r12}
    Determining the winner of a Maker-Breaker positional game is \PSPACE-complete even restricted to hypergraph of max degree $5$ and rank $12$.
\end{theorem}

\begin{proof}
    We provide a reduction from $6$-uniform Maker-Breaker games, which is already known to be \PSPACE-complete~\cite{rahman2021}. Let $\hxf$ be a hypergraph of rank~$6$. Let $\som = \{u_1, \dots, u_n\}$. We construct a hypergraph $\hxfp$ as follows:
\begin{enumerate}
    \item For each vertex $u_i \in \som$ we create four vertices $v_i^1$, $v_i^2$, $w_i^1$ and $w_i^2$ in $\som'$.
    \item For each hyperedge $e = \{u_1, u_2, \dots, u_6\} \in \WS$, we create the $2^6 = 64$ following hyperedges in $\WS'$: $\{v^{j_1}_1, w^{j_1}_1, v^{j_2}_2, w^{j_2}_2, \dots, v^{j_6}_6, w^{j_6}_6\}$ for $1 \le j_1, \dots j_6 \le 2$. Informally, these two steps consists in transforming each vertex of $\hyp$ into two pairs of vertices, then for each hyperedge $e$ of $\hyp$, replacing each vertex of $e$ by one of the pairs corresponding to $e$ gives a hyperedge of $\hyp'$.
    
    \item For each $1\le i \le n$ and each pair of vertices $(v_i^1, w_i^1)$, denote by $e_1, \dots, e_k$ the edges containing $(v_i^1,w_i^1)$. For each $1 \le j \le k$ we create two vertices $(v_i^{e_j}, w_i^{e_j})$. We then build a complete binary tree $T_i^1$ with $k$ leaves, rooted in some vertex $t^1_{i,0}$. We denote by $t^1_{i,j}$ the vertices of this tree. For each vertex $t^i_j$ of $T_i^1$, we add two vertices $(v_{i,j}^1, w_{i,j}^1)$. At the root, we consider $(v_i^1,w_i^1) = (v_{i,0}^1, w_{i,0}^1)$, and at the leaves, we consider for $1 \le p \le k$ $(v_{i,j_p}^1,w_{i,j_p}^1) = (v_i^{e_j}, w_{i}^{e_j})$.  We proceed similarly for $(v_i^2, w_i^2)$, creating a tree $T_i^2$.

    \item For each $1 \le i \le n$,  and each edge $(t^i_j, t^i_{j'})$, with $t^i_{j'}$ parent of $t^i_{j}$ of $T_i^1$, we add two vertices $a^1_{i,j}$ and $b^1_{i,j}$. Then, we add two hyperedges, $(v_{i,j'}^1, w_{i,j'}^1, v_{i,j}^1, a^1_{i,j})$ and $(v_{i,j'}^1, w_{i,j'}^1, w_{i,j}^1, b^1_{i,j})$. We proceed similarly with $T_i^2$.

    \item Finally, for each $1 \le i \le n$, we add five vertices $x_i$, $a^1_i$, $b^1_i$, $a^2_i$ and $b^2_i$ and the hyperedges $\{x_i, a^j_i, v^j_i\}$ and $\{x_i, b^j_i, w^j_i\}$ for $1 \le j \le 2$. We denote by $T_i$ all the vertices created for the same $i$ during the steps 1, 2 and 3. 
\end{enumerate}

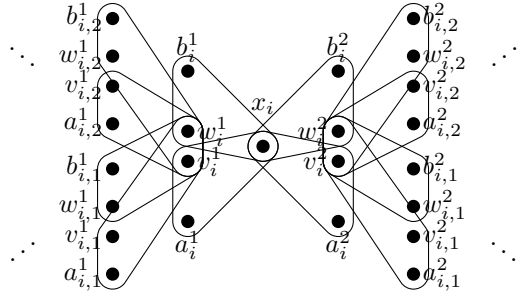
\begin{figure}
    \centering
\begin{tikzpicture}

\draw (0,0) node[v] (xi)  {} node[above = .3cm] {$x_i$};
\draw (-1,-1) node[v] (ai1)  {} node[below] {$a_i^1$};
\draw (-1,-.2) node[v] (vi1)  {} node[right] {$v_i^1$};

\draw (-1,1) node[v] (bi1)  {} node[above] {$b_i^1$};
\draw (-1,.2) node[v] (wi1)  {} node[right] {$w_i^1$};

\draw \hedgeiii{xi}{ai1}{vi1}{2mm};
\draw \hedgeiii{xi}{wi1}{bi1}{2mm};

\draw (-2,-1.7) node[v] (ai11) {} node[left] {$a_{i,1}^1$};
\draw (-2,-1.2) node[v] (vi11)  {} node[left] {$v_{i,1}^1$};

\draw (-2,-.3) node[v] (bi11)  {} node[left] {$b_{i,1}^1$};
\draw (-2,-.8) node[v] (wi11)  {} node[left] {$w_{i,1}^1$};

\draw \hedgeiiii{wi1}{vi1}{ai11}{vi11}{2mm};
\draw \hedgeiiii{wi1}{vi1}{wi11}{bi11}{2mm};

\draw (-2,. 3) node[v] (ai21) {} node[left] {$a_{i,2}^1$};
\draw (-2, .8) node[v] (vi21)  {} node[left] {$v_{i,2}^1$};

\draw (-2,1.7) node[v] (bi21)  {} node[left] {$b_{i,2}^1$};
\draw (-2,1.2) node[v] (wi21)  {} node[left] {$w_{i,2}^1$};

\draw \hedgeiiii{wi1}{vi1}{ai21}{vi21}{2mm};
\draw \hedgeiiii{wi1}{vi1}{wi21}{bi21}{2mm};

\draw (1,-1) node[v] (ai2)  {} node[below] {$a_i^2$};
\draw (1,-.2) node[v] (vi2)  {} node[left] {$v_i^2$};

\draw (1,1) node[v] (bi2)  {} node[above] {$b_i^2$};
\draw (1,.2) node[v] (wi2)  {} node[left ] {$w_i^2$};

\draw \hedgeiii{xi}{vi2}{ai2}{2mm};
\draw \hedgeiii{xi}{bi2}{wi2}{2mm};

\draw (2,-1.7) node[v] (ai12) {} node[right] {$a_{i,1}^2$};
\draw (2,-1.2) node[v] (vi12)  {} node[right] {$v_{i,1}^2$};

\draw (2,-.3) node[v] (bi12)  {} node[right] {$b_{i,1}^2$};
\draw (2,-.8) node[v] (wi12)  {} node[right] {$w_{i,1}^2$};

\draw \hedgeiiii{vi2}{wi2}{vi12}{ai12}{2mm};
\draw \hedgeiiii{vi2}{wi2}{bi12}{wi12}{2mm};

\draw (2,. 3) node[v] (ai22) {} node[right] {$a_{i,2}^2$};
\draw (2, .8) node[v] (vi22)  {} node[right] {$v_{i,2}^2$};

\draw (2,1.7) node[v] (bi22)  {} node[right] {$b_{i,2}^2$};
\draw (2,1.2) node[v] (wi22)  {} node[right] {$w_{i,2}^2$};

\draw \hedgeiiii{vi2}{wi2}{vi22}{ai22}{2mm};
\draw \hedgeiiii{vi2}{wi2}{bi22}{wi22}{2mm};

\draw (3.2,1.3) node {$\iddots$};
\draw (-3.2,1.3) node {$\ddots$};

\draw (-3.2,-1.3) node {$\iddots$};
\draw (3.2,-1.3) node {$\ddots$};

\end{tikzpicture}    

\caption{Gadget for the vertices in Maker-Breaker}
    \label{fig: MB bounded degree}
\end{figure}

Note that this construction is polynomial since $|\som'| = O(|\som|^2*|\WS)$, $\hyp'$ has rank $12$, and $\Delta(\hyp') = 5$. It remains to prove that $o(\hyp) = o(\hyp')$.

Overview of the construction: a vertex $x_i$ corresponds to the vertex $u_i \in \som$. If Maker plays it, she can always claim all the vertices $v_{i,j}^\varepsilon$ and $w_{i,j}^\varepsilon$ for $\varepsilon = 1$ or $\varepsilon = 2$, ensuring her to claim the corresponding vertices in the hyperedges where $u_i$ appears, since any of these moves force a move from Breaker. If Breaker claims it, he can always pair all the $(v_{i,j}^\varepsilon, w_{i,j}^{\varepsilon}$ for $\varepsilon \in \{1,2\}$, ensuring him to claim at least one vertex in each hyperedge created from a hyperedge where $u_i$ is in $\hyp$.

For sake of simplicity in the notation, we will say that a player claims a vertex in a tree $T$, to say that he or she claims a vertex that was added to $\hyp'$ during the step considering $T$ in the construction.

Suppose first that Breaker has a winning strategy $\strat$ on $\hyp$. We propose the following strategy for him on $\hyp'$. 
\begin{itemize}
    \item If Breaker starts, if his first move in $\hyp$ is a vertex $u_i$, then he starts by playing $x_i$.
    \item If Maker plays for the first time a vertex in $T_i^1 \cup T_i^2$ for $\varepsilon \in \{1,2\}$, Breaker considers that she has played $u_i$ in $\hyp$, and plays the vertex $x_j$ corresponding to the vertex $u_j$ he would have played as an answer to $u_i \in \strat$.
    \item If Maker plays a vertex in $T_i^1 \cup T_i^2$ with an already played vertex in $T_i^1 \cup T_i^2$, Breaker answers according to the following pairing strategy: \begin{itemize}
        \item If Breaker claimed the first vertex in $T_i$, necessarily, it was $x_i$. The Breaker consider the pairs $\{v_{i,j}^\varepsilon, w_{i,j}^\varepsilon\}$ for $1 \le i \le n$, $1 \le j \le d(u_i)$, and $\varepsilon \in \{1,2\}$.
        \item If Maker claimed the first vertex $x$ in $T_i^1 \cup T_i^2$, suppose without loss of generality that it was in $T_i^1$. Breaker considers the pairs $\{v_{i,j}^\varepsilon, a_{i,j}^\varepsilon\}$ and $\{v_{i,j}^\varepsilon, a_{i,j}^\varepsilon\}$. If $x = x_i$, this pairing does not need to be modified. If $a \neq x_i$, since any hyperedge containing $a_{i,j}^\varepsilon$ ( resp. $b_{i,j}^\varepsilon$) also contains $v_{i,j}^\varepsilon$ (resp. $w_{i,j}^\varepsilon$), we can suppose that $x = v_{i,j}^\varepsilon$ or $x = w_{i,j}^\varepsilon$. By symmetry, suppose $x = v_{i,j}^\varepsilon$. Let $t^{\varepsilon}_{i,j_0} = t^{\varepsilon}_{i,0}, \dots, t^{\varepsilon}_{i, j_p}$ be a path from the root to the vertex corresponding to $x$ in $T_i^\varepsilon$. Breaker, along this path, changes the pairing by pairing $(a^\varepsilon_{i,j_l}, v^\varepsilon_{i,j_{l-1}}$, for $2 \le l \le p$, and $(a^\varepsilon_{i,j_1}, x_i)$. Note that this change is possible since now, $x_i$ is unclaimed.
    \end{itemize}
\end{itemize}

By construction, any hyperedge containing only vertices of some tree $T^\varepsilon_i$ contains a pair of paired vertices from Breaker. Therefore, Breaker can only lose on hyperedges related to the hyperedges creating during step~2 of the construction. Let $e$ be one of these hyperedges, and consider $e_i$ the hyperedge of $\hyp$ that was consider to create $e$. Since, $\strat$ was a winning strategy for Breaker in $\hyp$, there exists $u_i \in e$ such that Breaker has played $u_i$ according to strat. Thus, he has played $x_i$ according to $\strat'$, and finally, he has paired all the pairs $(v^\varepsilon_{i,j}, w^\varepsilon_{i,j}$ in $T^\varepsilon_i$. Since, by construction, one of these pairs is in $e$, Breaker has claimed at least one vertex of $e$. Thus Breaker wins.  

Suppose now that Maker has a winning strategy $\strat$ on $\hyp$. We propose the following strategy for him on $\hyp'$.

\begin{itemize}
    \item If Maker starts, if her first move in $\hyp$ is a vertex $u_i$, then she starts by playing $x_i$.
    \item If Breaker plays a first vertex $x_i$ in $T_i$ while the other vertices in $T_i$ are unclaimed, Maker suppose that he has played $u_i \in \hyp$ and plays the vertex $x_j$ corresponding to the vertex $u_j$ she would have played according to $\strat$.
    \item If Breaker plays a vertex $y$ in $T_i$, Maker answers depending on the player who played the first move in $T_i$. If it was Breaker, Maker plays an arbitrary vertex of $T_i$. If Maker played the first vertex in $T_i$, by construction of $\strat'$, he played $x_i$. Thus $y$ is either in $T_i^1$ or in $T_i^2$. Suppose, without loss of generality, that $y \in T_i^2$. Maker plays, from the root to the leaves of $T_i^1$ the vertices $v^1_{ij}$ and $w_{i,j}^1$ in that order, forcing answers of Breaker on $a_{i,j}^1$ and $b_{i,j}^1$. If at some points Breaker does not play this forced move, Maker claims it and wins instantly.
    \item Once all the vertices $x_i$ are claimed, if Maker has not won yet, for each vertex $x_i$ he has claimed, he claims the vertices of $T_i^1$ following the above point.
\end{itemize}

Following this strategy, if Maker has not won in a Tree $T_i$, for each vertex $x_i$ he has claimed, he has either claimed all the $(v_{i,j}^1, w_{i,j}^1)$ or all the $(v_{i,j}^2, w_{i,j}^2)$. Since $\strat$ is a winning strategy in $\hyp$, there exists a hyperedge $e \in \WS$ such that Maker claims all the vertices of $e$ following $\strat$. Let $u_{i_1}, \dots, u_{i_6}$ be the vertices of $e$. By construction of $\strat'$, Maker has played $x_{i_1}, \dots, x_{i_6}$ in $\hyp'$. Thus, there exists $1\le \varepsilon_1, \dots, \varepsilon_6 \le 2 $ such that Maker has claimed all the $v^{\varepsilon_k}_{i_k,j}$ and all the $w^{\varepsilon_k}_{i_k,j}$ for $1 \le k \le 6$. Thus by construction, she has filled up one of the hyperedges created during step~2 and she won.

Finally, $o(\hyp) = o(\hyp')$, which concludes the reduction.
\end{proof}

As a consequence, looking at the \PSPACE-completeness reduction provided by Duchêne \etal in~\cite{Duchene2020} for the Maker-Breaker Domination game in the bipartite case, the same reduction proves that the Maker-Breaker domination game is \PSPACE-complete, even restricted to graphs of maximum degree $24$, which answers an open question of Bagan \etal~\cite{Bagan2024} asking if some positive results could be obtained for bounded degree graphs in the domination game.

\begin{corollary}
    Determining the winner in the Maker-Breaker Domination game is \PSPACE-complete, even restricted to graph of maximum degree $24$.
\end{corollary}

\begin{proof}
    This result is straightforward since the construction of Duchêne \etal~\cite{Duchene2020} creates a graph with one vertex per vertex of the hypergraph, two vertices per hyperedges of the hypergraphs and connects two vertices if they corresponds to a vertex and a hyperedge containing it in the hypergraph.
\end{proof}

\subsection{Maker-Maker games}

We recall that in Maker-Maker convention, two players, Alice and Bob, alternate claiming the vertices of a hypergraph. If at some point, one player manage to fill up a hyperedge, he wins. If it doesn't happen and there is no unclaimed vertex left, the outcome is a Draw.

In 2004, Byskov~\cite{byskov2004} proved that computing the winner of a Maker-Maker games is \PSPACE-complete, by reduction from Maker-Breaker adding a universal vertex. Although we cannot add a universal vertex without removing the bound on the maximum degree of the hypergraph, a similar construction forcing all the moves of the second player can be used to prove that Maker-Maker games are also \PSPACE-complete for bounded degree graphs.

\begin{theorem}
    Determining the winner of a Maker-Maker positional game is \PSPACE-complete even restricted to hypergraph of max degree $5$ and rank $13$.
\end{theorem}

\begin{proof}
    The proof is a reduction from Maker-Breaker games on rank~$12$ and maximum degree~$5$ hypergraphs. Let $\hxf$ be a hypergraph. We construct a hypergraph $\hxfp$ as follows:

\begin{itemize}
    \item We start with a copy of $\hyp$, i.e. $\som' = \som$ and $\WS' = \WS$.
    \item Denote by $e_1, \dots, e_m$ the hyperedges of $\hyp$. For each hyperedge $e_i \in \WS$, we add two vertices $x_i, y_i$ to $\som'$. We add $x_i$ into $e_i$ and we create a new hyperedge $\{x_i, y_i\}$.  
\end{itemize}

We prove that Maker wins in $\hyp$ if and only if Alice wins in $\hyp'$.

Suppose first that Maker has a winning strategy $\strat$ in $\hyp$, and consider the following strategy for Alice in $\hyp'$:
\begin{itemize}
    \item For $i = 1$ to $m$, Alice plays $x_i$. If Bob does not answers by playing $y_i$, Alice plays it instantly and wins.
    \item If she has not won during the first step, all the vertices $x_i$ and $y_i$ have been played. Now, she can play following $\strat$ since she has already played a vertex in all the hyperedges and thus Bob cannot win.
\end{itemize}

Reciprocally, if Breaker has a winning strategy $\strat$ in $\hyp$, Bob can play as follows:
\begin{itemize}
    \item If Alice claims a vertex in some pair $\{x_i, y_i\}$, he answers by claiming the second vertex of the same pair.
    \item Otherwise, Alice has played a vertex which also exists in $\hyp$ and Bob plays according to $\strat$.
\end{itemize}

Following this strategy, Bob plays according to $\strat$ in $\hyp$, thus, since $\strat$ is a winning strategy for Breaker, Alice cannot win in any hyperedge $e_i$. In all the other hyperedges, Bob has played either $x_i$ or $y_i$, and thus Alice has no filled up the hyperedge. Finally, this strategy ensures at least a draw for Bob, which ends the proof, since the second player cannot have a winning strategy in a Maker-Maker game.
\end{proof}

\bibliography{biblio}
\bibliographystyle{alpha}

\end{document}